\documentclass{PoS}

\usepackage{cite}

\def\ttt#1{\texttt{\small #1}}

\providecommand{\gaga}{\gamma\,\gamma}
\providecommand{\epem}{\rm e^{+}e^{-}}
\providecommand{\alphas}{\alpha_{\rm s}}

\newcommand{\sqrts}{\sqrt{s}}
\newcommand{\ttbar}{t\overline{t}}

\title{QCD and $\gaga$ studies at the FCC-ee}

\ShortTitle{QCD and $\gaga$ Studies at FCC-ee}

\author{\speaker{Peter Skands}\\
School of Physics and Astronomy, Monash University, VIC-3800, Australia\\
        E-mail: \email{peter.skands@monash.edu}}

\author{David d'Enterria\\
        CERN, EP Department, CH-1211 Geneva 23, Switzerland\\
        E-mail: \email{dde@cern.ch}}

\abstract{The Future Circular Collider (FCC) is a post-LHC project aiming at searches for physics beyond 
the SM in a new 80--100~km tunnel at CERN. Running in its first phase as a very-high-luminosity 
electron-positron collider (FCC-ee), it will provide unique possibilities for indirect searches of 
new phenomena through high-precision tests of the SM. In addition, by collecting tens of ab$^{-1}$ integrated 
luminosity in the range of center-of-mass energies $\sqrts$~=90--350~GeV, the FCC-ee also offers 
unique physics opportunities for precise measurements of QCD phenomena and of photon-photon collisions 
through, literally, billions of hadronic final states as well as
unprecedented large 
fluxes of quasireal $\gamma$'s radiated from the $\epem$ beams.
We succinctly summarize the FCC-ee perspectives for high-precision
extractions of the QCD coupling, for detailed analyses of parton radiation and fragmentation,
and for SM and BSM studies through $\gaga$ collisions.} 

\FullConference{38th International Conference on High Energy Physics\\
		3-10 August 2016\\
		Chicago, USA}

\begin{document}

\section{Introduction}

Electron-positron collisions at the FCC-ee not only provide 
unparalleled opportunities for electroweak and beyond standard model (BSM) physics~\cite{Gomez-Ceballos:2013zzn}, 
but also offer a vast landscape of possibilities for 
precise measurements of QCD phenomena and of photon-initiated processes. Beyond their intrinsic value, such 
measurements will also leave an important legacy for any subsequent hadron collider, much as that from LEP 
and other $\epem$ colliders provided crucial benchmark constraints e.g.\ on the Monte Carlo parton radiation/fragmentation 
models used at the LHC. The FCC-ee working group WG5\footnote{Participation in FCC-ee WG5 is open to anyone 
interested in studies of QCD and/or $\gaga$ physics via the main
FCC-ee site: \ttt{http://CERN.ch/FCC-ee} (join us, subscribe).}  aims at quantitatively exploring all such potentialities, with
a view to informing a CERN Yellow Report on FCC-ee physics in 2017. 
Briefly summarized, these are the WG5 physics goals:
\begin{enumerate}
\item 
Determine the best achievable experimental and theoretical precision on the extraction of the QCD coupling $\alpha_s$
  ~\cite{d'Enterria:2015toz,d'Enterria:2016zpn}.
\item Exploit the unique high-precision QCD physics opportunities in the clean $\epem$ environment,
  with a view to improving future studies at pp colliders, including
  studies of QCD multijets, jet substructure, quark-gluon discrimination, q,g,c,b,(t) parton-to-hadron
  fragmentations, colour reconnections, multiparticle
  correlations, and rare hadron production and decays. 
\item Assess photon-photon ($\gaga$) physics possibilities for SM and BSM studies.
\item Set goals for subdetector performance (including forward
  e$^\pm$ taggers for $\gaga$ physics, particle identification
  for hadronization studies, jet resolution requirements for precision
  QCD, etc.) and experimental conditions so that systematic
  uncertainties are of similar magnitude as statistical uncertainties.
  \item Define experimental/phenomenological software needs to enable
  the measurements and precision interpretations. 
\item Help evaluating the QCD impact on the rest of the FCC-ee physics
  program. In particular, establish background event generators for QCD and
  $\gaga$ processes. 
\end{enumerate}

\begin{figure}[t]
\centering
\includegraphics[scale=0.3]{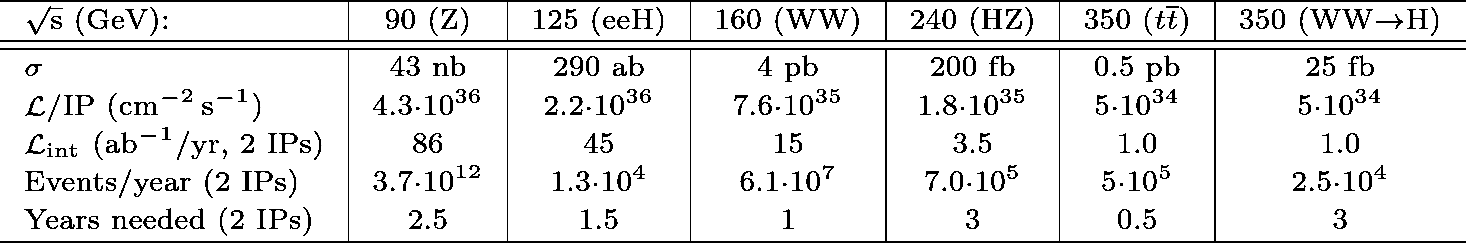}
\includegraphics[scale=1.25]{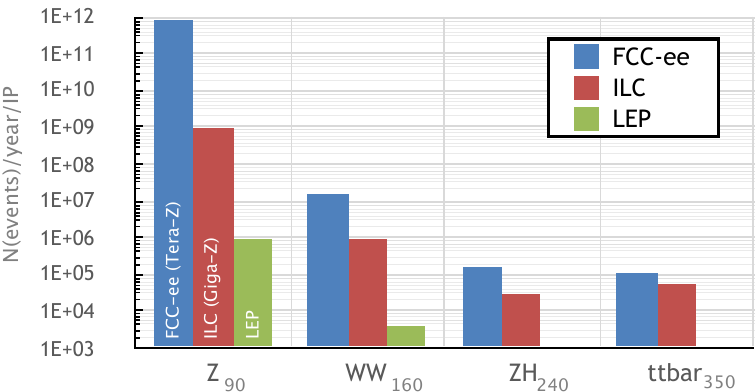}
\caption{Top: Target luminosities, events/year, and years needed to complete the W, Z, H and
   top programs at FCC-ee. [$\cal{L}$~=~10$^{35}$~cm$^{-2}$\,s$^{-1}$ corresponds to
$\cal{L}_{\rm int}$~=~1~ab$^{-1}$/yr for 1 yr = 10$^7$~s]~\cite{d'Enterria:2016yqx}.
Bottom: Expected number of events/year, per interaction point, 
at the FCC-ee 
compared with ILC(Giga-Z) and LEP. 
\label{fig:numbers}}
\end{figure}

With the exception of $\alpha_s$ determinations, the target studies
are only just beginning, hence the points below are intended
mainly as exhortations, highlighting some of the exciting possibilities. 
To set the stage, Table~\ref{fig:numbers} lists the expected event
numbers per year and interaction point (IP), for Z, W\,W, Z\,H, and
$\ttbar$ at the FCC-ee (compared with ILC~\cite{ILC} and LEP in the bottom panel).
In addition to accessing the previously unreachable Z\,H and $t\bar{t}$ thresholds, both 
ILC and FCC-ee clearly offer vastly increased statistics with respect to LEP, 
not only for Z but, in particular, the available 
samples of W\,W events will increase from about 11\,000 (per IP) integrated 
over the full LEP2 running period to tens of millions, enabling truly
high-statistics $\rm \epem\to W^+W^-$ measurements for the first
time. This will be a highly fruitful testing ground, e.g.\ for
colour reconnection studies (likewise for $t\bar{t}$ events), see e.g.\ \cite{Christiansen:2015yca},
and for precision W-based extractions of $\alphas$, competitive with the determinations at the Z pole.

\section{High-precision $\alphas$ determination}

\begin{figure}[htbp!]
  \centering
   \includegraphics[width=0.465\columnwidth]{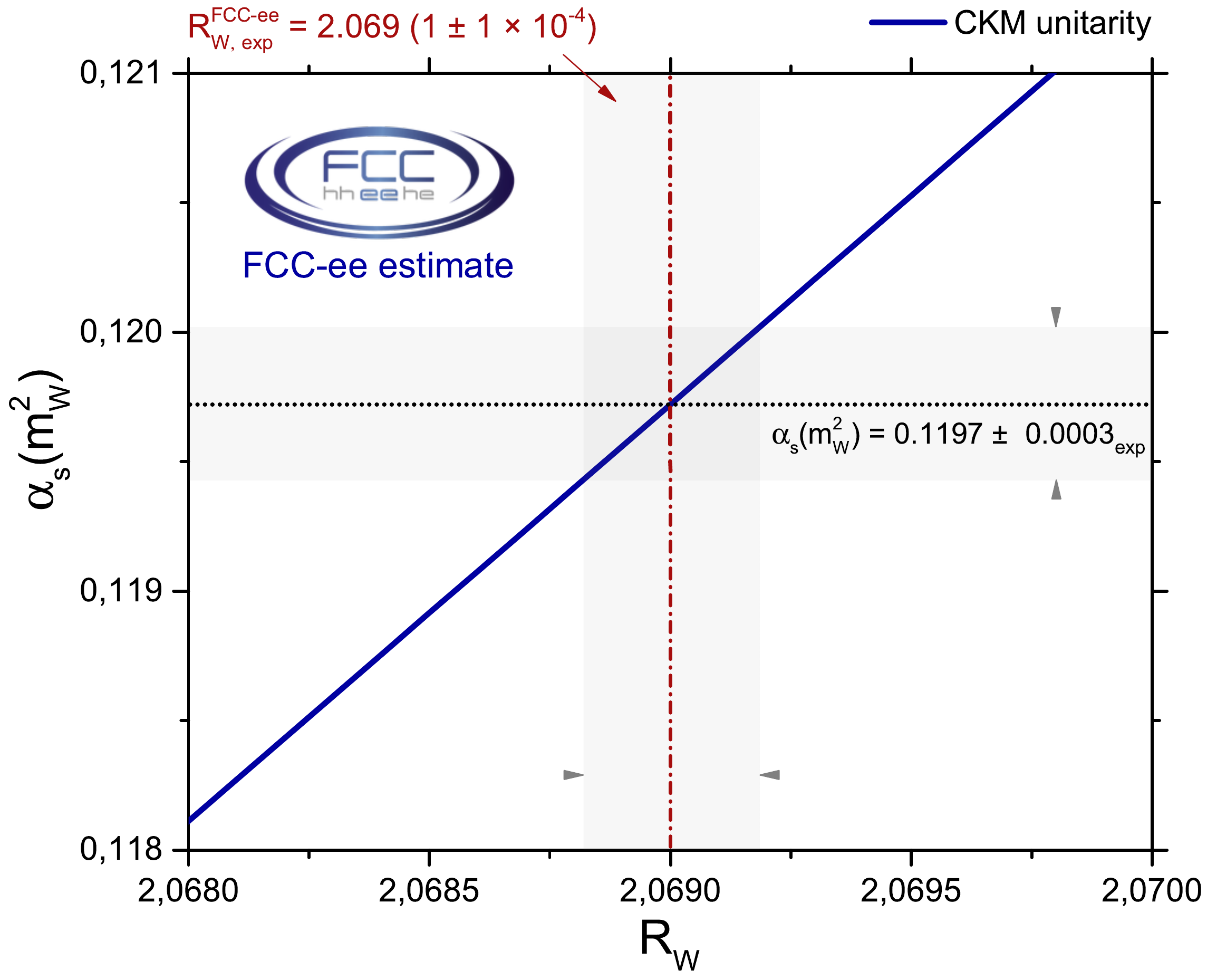}
   \includegraphics[width=0.525\columnwidth]{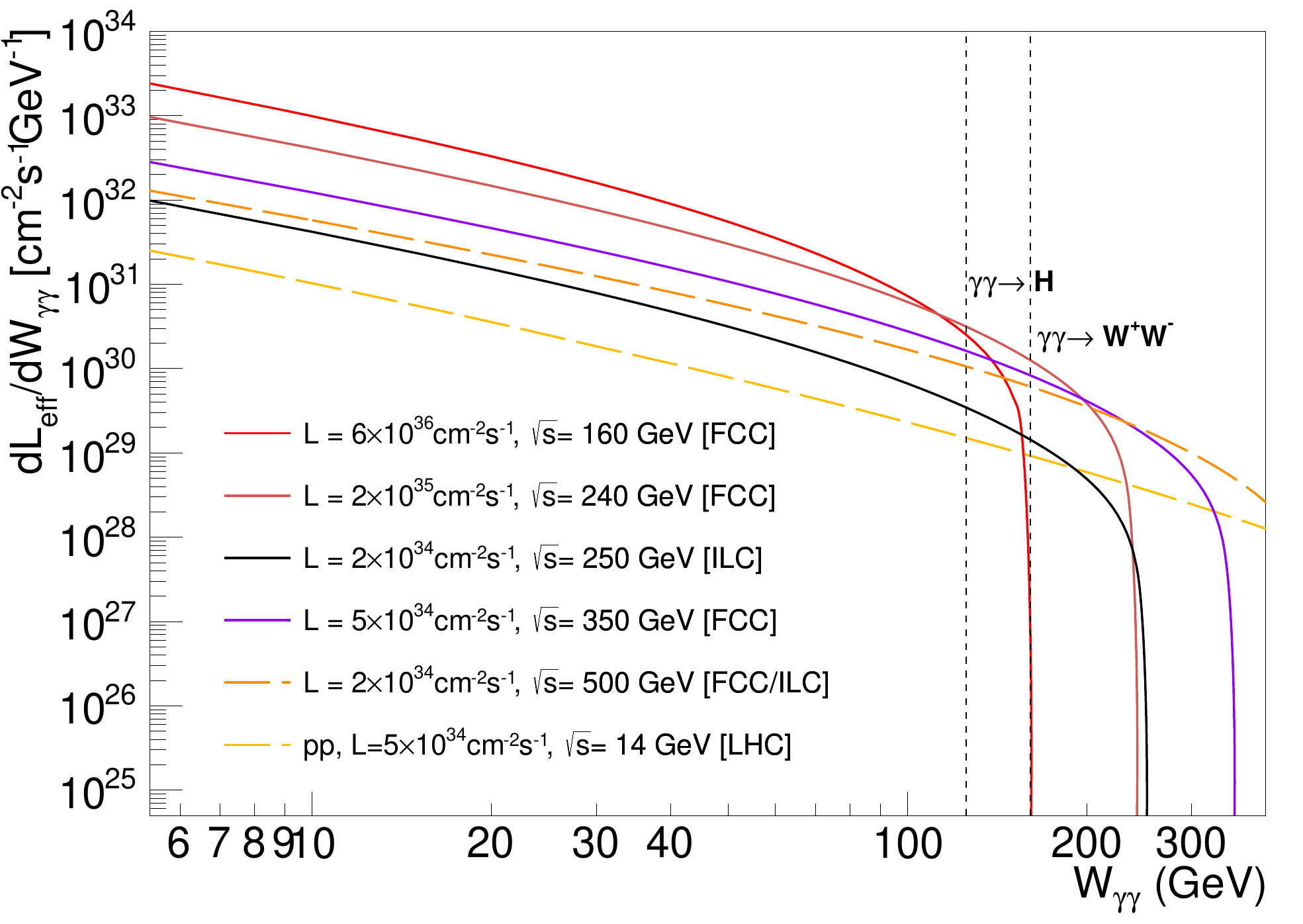}
  \caption{Left: Extraction of $\alphas$ from the hadronic/leptonic W decay ratio ($\rm R_{W}$) 
  expected at the FCC-ee~\cite{d'Enterria:2016ujp}.
  Right: Effective two-photon luminosities as a function of $\gaga$ CM energy
  at FCC-ee, ILC, 
  and pp at the LHC, for their nominal beam luminosities (vertical lines show 
  $\rm \gaga \to H, W^+W^-$ production thresholds)~\cite{Teles:2015xua}.}
\label{fig:alphas_gaga}
\end{figure}

The combination of numerous high-precision hadronic observables will lead to an $\alphas$ 
determination with permille uncertainty at the FCC-ee as discussed in~\cite{d'Enterria:2015toz,d'Enterria:2016zpn}.
First, the huge statistics of hadronic $\tau$, W, and Z decays,  
studied with state-of-the-art perturbative calculations, will provide $\alphas$ extractions 
with very small uncertainties: $<1$\% from $\tau$, and $<$0.2\% from W and Z bosons.
Fig.~\ref{fig:alphas_gaga} left shows the expected result from the
ratio of W hadronic/leptonic decays ($\rm R_{W}$) alone~\cite{d'Enterria:2016ujp}.
In addition, the availability of millions of jets (billions at the Z pole) 
measured over a wide $\sqrts\approx$~90--350~GeV range, with light-quark/gluon/heavy-quark discrimination and reduced hadronization
uncertainties (whose impact decreases roughly as 1/$\sqrts$), will provide $\alphas$ extractions with 
$<$1\% precision from various independent observables: hard and soft fragmentation functions, jet rates, and event shapes.
Last but not least, $\gaga\to$~hadrons collisions will allow for an accurate extraction of the QCD photon structure
function ($F_2^\gamma$) and thereby of $\alphas$. 

\section{High-Precision QCD studies}

The extremely large event numbers for hadronic Z decays also make it possible to
conceive very precise measurements of multijets (potentially applying aggressive cuts to probe
topologies of specific interest, such as ``hedgehogs'' sensitive to
multi-partonic coherence effects~\cite{Fischer:2014bja}), 
gluon- and $b$-jet fragmentation (including $g\to b\bar{b}$
splittings), and colour-reconnection effects inside Z 
decays. Importantly, good particle-identification
capabilities would open up a whole world of detailed
measurements to constrain the nature of hadronization, shining light
on the confinement mechanism in QCD, and its dependence on quark
and hadron masses and quantum numbers, with better statistics and
reduced systematics compared with equivalent measurements from earlier
$\epem$ colliders. This includes not 
only single-particle spectra and multiplicities, but also measurements
of Bose-Einstein, Fermi-Dirac, rapidity, spin, and flavour
correlations. Proof-of-concept phenomenology studies are needed in all
these areas.  
Searches can also be undertaken for 
the production of very rare hadrons which require the ``coalescence''
of multiple perturbatively created quarks, such as double- or
triple-$(c,b)$ baryons, in addition to the more traditionally studied
onia states, the latter of which can be created from a single $g\to Q\bar{Q}$
splitting followed by colour reconnection.\\

Especially for jet rates and jet substructure measurements, it will be
crucial to establish how the improvements in detector resolution, in
conjunction with modern analysis techniques, will impact the achievable
systematic uncertainties.  For a representative value of Durham $k_T
\sim 4.8\ \mathrm{GeV}$ (large enough to effectively suppress hadronization
corrections but small enough to allow significant event numbers), already the
LEP statistics of order 1M hadronic Z decay events yield a statistical precision
on the total jet rates below the 1\% level for up to 6 jets, see Table~\ref{tab:jets}. 

\begin{table}[hbpt]
\centering
\begin{tabular}{r|r|r|r|r}
$N_J$ & 3 & 4 & 5 & 6 \\\hline
$\ge N_J$ (incl) & 640k & 240k & 60k & 10k \\
 $N_J$ (excl) & 400k & 180k & 50k & 
\end{tabular}
\caption{Number of $N_J$-jet events 
  per 1M hadronic Z decays, for $\ln(y_\mathrm{cut}) =
  \ln(k_T^2/E_\mathrm{CM}^2) = -5.9$. \label{tab:jets}}
\end{table}
However, the systematic uncertainties at LEP were roughly an order of
magnitude larger, with the existing measurements typically
exhibiting a 10\% or larger total uncertainty~\cite{Heister:2003aj}. This 
is already now becoming insufficient to discriminate between the current state-of-the-art
calculations and MC models~\cite{Karneyeu:2013aha}, 
let alone those that will be around a few decades from now. 
Better calorimetric resolutions and use of
more advanced analysis techniques, such as particle flow, are expected
to make it possible to reduce these errors substantially. The
larger statistics will allow probing higher $k_T$ values as well as
placing cuts to focus on specific topologies within these
samples, and measure e.g.\ substructure- or coherence-sensitive
observables with high precision, revealing details of quark \& gluon
jet (sub)structure beyond the (N)LL level and the patterns in which
multiple soft (but perturbative) gluons are emitted from a system of
$N$ hard partons.

\section{Photon-photon physics}

Photon-fusion processes 
can also be studied exploiting the large flux of quasireal 
photons radiated from the $\epem$ beams, theoretically 
described using the effective photon approximation. 
The $\gaga$ kinematics can be constrained measuring the scattered $\rm e^\pm$ with near-beam detectors, 
while the produced system is reconstructed from its decay products in the central detector.
For $\gaga$ collisions, the available e$^\pm$ beam luminosity is reduced but
nonetheless still substantial (Fig.~\ref{fig:alphas_gaga}, right). 
The $\gaga$ luminosity with more than 10\% (50\%) of the full CM energy is estimated to be
 about 1\% (0.04\%) of the $\epem$ one, allowing for precise studies of hadronic, W\,W,
$\gaga$, and $\tau\tau$ final states, as well as about 100 $\rm \gaga\to H$ events per
$\mathrm{ab}^{-1}$~\cite{Teles:2015xua}. In the QCD sector, since $\sigma(\gaga \to
\mathrm{hadrons}) \propto 1/s$, at high $\sqrt{s}$ the rate of $\gaga$-induced
hadronic events can actually be higher than the latter despite the factor
$10^2$--$10^3$ luminosity penalty, and the $t$-channel nature of the
process may open an interesting window on BFKL-type ladders. In the electroweak sector,
the measurement of $\rm \gaga \to W\,W \to 4$~jets will yield more than 600 final counts 
which will allow for detailed studies of the trilinear W\,W$\gamma$ and quartic W\,W\,$\gaga$
couplings, either in the SM or assuming new physics scenarios in terms of dimension-6 and 8 effective
operators~\cite{Teles:2015xua}. 

\section{Conclusion}
A future  $e^+e^-$ collider operating in the energy range between 90--350 GeV
centre-of-mass energy provides physics opportunities for QCD and
$\gamma\gamma$ studies that go well beyond what was achievable at
earlier machines. The expected improvements in detector
performance, resolution, and analysis methods, will form a powerful
combination with vastly increased statistics for the $Z$ and, especially, W\,W channels, access to the Z\,H and $t\bar{t}$
thresholds for the first time, and huge $\gamma\gamma$
luminosities. In this contribution, we summarised some of the key
numbers relevant to the FCC-ee and pointed out areas which could benefit from further
studies. While the potential for precision $\alpha_s$ extractions is
already reasonably well
explored~\cite{d'Enterria:2015toz,d'Enterria:2016zpn}, we encourage
the community to consider seriously the new opportunities that are opened in the
areas of precision QCD jets and fragmentation studies, 
and for $\gamma\gamma$ collisions. The former will be the topic of an
upcoming workshop at CERN, Nov 21--22 2016
(\texttt{https://indico.cern.ch/e/ee\_jets16}) while a recent discussion
of the latter can be found in~\cite{Teles:2015xua}.

\paragraph{Acknowledgments:} PS is the recipient of an Australian Research Council Future Fellowship, 
FT130100744: ``Virtual Colliders: high-accuracy models for high energy physics''.

\bibliographystyle{JHEP}
\bibliography{ichep}

\end{document}